# Semantic Interoperability on Blockchain by Generating Smart Contracts Based on Knowledge Graphs

William Van Woensel[a] and Oshani Seneviratne[b]

[a]Telfer School of Management, University of Ottawa, 55 Laurier E., Ottawa ON K1N 6N5, Canada
[b]Rensselaer Polytechnic Institute, Troy NY 12180, USA

## Abstract

**Background**: Health 3.0 marks the next step in the evolution of healthcare, where data management moves from being institution-centric and centralized to patient-centric and distributed. Health 3.0 thus enables decision making to be based on longitudinal data from multiple institutions, from across the patient's healthcare journey. In such a distributed setting, blockchain can act as a neutral intermediary to implement trustworthy decision making. Blockchain, as a separate, secure and transparent platform, does not require any party to entrust another with their privacy-sensitive data, nor to execute the decision-making process. Instead, secure transactions share patient data with a smart contract, which implements the decision-making process on blockchain in a transparent and tamper-proof way. Therefore, contributing parties only require trust in the original smart contract and the underlying blockchain technologies (consensus mechanism, ledger).

**Objective**: In a distributed setting, transmitted data will be structured using standards (such as HL7 FHIR) for semantic interoperability. In turn, the smart contract will require interoperability with the domain standard; implement a complex communication setup to work in a distributed environment (e.g., using oracles); and be developed using special-purpose blockchain languages (e.g., Solidity). We propose the encoding of smart contract logic using a high-level semantic Knowledge Graph (KG), using concepts and relations from a domain standard, and describing data requirements from distributed parties. We subsequently deploy this semantic KG on blockchain for trustworthy decision making.

**Methods**: We use a hybrid on-/off-chain solution to deploy the semantic KG on blockchain: off-chain, a code generation pipeline compiles the KG into a concrete smart contract, which is then deployed on-chain. Our pipeline targets an intermediary "bridge" representation, which captures declarative logic using imperative constructs, and is then "transpiled" into a specific blockchain language. Our choice for off-chain code generation avoids on-chain rule engines, with unpredictable and likely higher computational cost. It is thus in line with the economic rules of blockchain, where heavier computations result in higher execution costs.

**Results**: We applied our code generation approach to generate smart contracts for 3 health insurance cases from Medicare. We discuss the suitability of our approach—the need for a neutral intermediary—for a number of healthcare use cases. We evaluated the generated contracts in terms of correctness and execution cost ("gas") on blockchain, finding that they perform well in comparison.

**Conclusions**: We showed that it is feasible to automatically generate smart contract code based on a semantic KG, in a way that respects the economic rules of blockchain. Future work includes supporting a more expressive set of N3 rules, studying the use of Large Language Models (LLM) in supporting our approach, and evaluations on other blockchains.

# Introduction

Health 3.0 [1] marks a paradigm shift in healthcare data and service management, away from an institution-centric model with one-size-fits-all digital services and centralized closed-off data silos. Instead, Health 3.0 is characterized by a *patient-centric model*, where healthcare services are personalized and customized to individual patient profiles and needs. Moreover, Health 3.0 aims to make a patient's longitudinal data, i.e., recording their healthcare journey across multiple institutions, accessible to the patient and their care providers for improved decision making. Health 3.0 thus implies a distributed environment, i.e., in which data will be integrated from multiple institutions. To implement health decision making in such a distributed environment, a suitable platform will be needed.

We argue that the unique characteristics of blockchain technology are conducive to this role. Blockchain can deploy decision making in a distributed setting as a *neutral intermediary*, i.e., without any party having to entrust another party with (a) their privacy-sensitive health data, or (b) executing the particular decision making process. At its core, blockchain is a secure and transparent third-party platform for conducting "transactions"; a transaction involves the exchange of data or currency, is validated by a consensus mechanism, and is recorded into an immutable ledger that is accessible to all parties. Smart contracts are blockchain programs for implementing decision making that involves multiple parties and transactions. When using a smart contract to implement health decision making, parties thus do not share health data directly with each other, but rather via secure transactions[1] with the smart contract on a separate platform. Moreover, after its publication on blockchain, a smart contract cannot be modified, its execution cannot be influenced or tampered with, and each execution is similarly recorded in the ledger. Hence, blockchain, and smart contracts, by extension, are "*trustless*": a contributing party does not need to place trust in another party, or any single provider, to run transactions and smart contracts; they only need to trust the original smart contract and proven blockchain technologies (e.g., consensus mechanism, ledger). Until now, smart contracts have mostly been applied to coordinate legal and financial interactions, however they have also been used for healthcare cases that require a neutral intermediary. Examples include ensuring that trial recruitment adheres to regulations [4]–[7], deciding on unbiased funding for prevention programs [8], and selecting health insurance and coverage [9]–[11].

---
[1] Note that transaction data can be encrypted to handle sensitive data [2], [3].

To implement decision making in any distributed environment, i.e., where data is contributed by multiple parties, transmitted data will need to be semantically interoperable. This can be achieved by structuring the data using well-known domain standards, such as HL7 FHIR [12]. When processing standards-based data, the smart contract code will thus need to be interoperable with the domain abstractions (e.g., "resources" in HL7 FHIR). This can be achieved by directly representing domain abstractions using internal Abstract Data Types (ADT), instantiating ADTs based on incoming data, and then processing them to implement the decision logic. Unfortunately, this is where we run into a contradiction: domain experts, i.e., with knowledge of domain standards, will be best positioned to write application logic in line with these standards [13]; but web 3.0 (i.e., smart contract) developers are most suited to implement complex smart contract code for a distributed Health 3.0 setting (e.g., requiring oracles for off-chain communication [14]). Development of complex smart contracts, using a tailored blockchain language (e.g., Solidity [15]), is typically beyond the purview of domain experts; vice-versa, web 3.0 developers tend to lack in-depth knowledge on comprehensive domain standards (e.g., HL7 FHIR).

We propose a declarative approach for domain experts to encode high-level smart contract logic, in terms of (a) high-level concepts and relations from a domain standard, and (b) high-level data requirements from distributed parties. To that end, we rely on Notation3 (N3) [16] as a declarative rule language, and OWL2 [17] for encoding a domain standard as an ontology. Using these two building blocks, domain experts can author *a semantic Knowledge Graph (KG) [18] that represents high-level smart contract logic in a distributed setting*. This KG will be deployed on blockchain using a hybrid on-/off-chain solution: off-chain, a *graph-based code generation pipeline* compiles the KG into a smart contract with imperative code, Abstract Data Types (ADT), and remote service invocations; then, the generated smart contract is deployed on-chain. Our choice for off-chain code generation is in line with the economic rules of blockchain-based systems, where higher computational workloads result in higher execution costs; we thus avoid on-chain rule engines with unpredictable and likely higher costs[2] [13]. To support multiple blockchain platforms (e.g., Solidity for Ethereum & Hyperledger[3], JavaScript for Hyperledger), our pipeline further targets a *"bridge" representation*, which captures declarative logic using general imperative programming constructs, and is then "transpiled" into specific blockchain languages.

We position our approach within the state-of-the-art in the Discussion section. We evaluate our approach from (a) an applicability viewpoint, with a look at how healthcare scenarios can benefit from blockchain deployment (similar to Choudhury et al. [19]), and (b) a software viewpoint, performing code analysis, validating correctness, and measuring the contracts' execution ("gas") cost (as by Karmakar et al. [9]). This paper expands upon our prior work [8] with applicability and performance evaluations, an extension for dynamic service invocation, and a more comprehensive description of the code generation pipeline, including pseudocode and discussing support for multi-valued properties and unification.

---

[2] We discuss this in more detail in the Discussion section.
[3] https://www.hyperledger.org/

# Methods

## Graph-Based Code Generation from Knowledge Graphs

We developed a graph-based approach for generating smart contracts from semantic Knowledge Graphs (KG), which include high-level declarative logic (N3 rules) that encodes policies and regulations, using domain concepts and relations from an OWL2 ontology. Figure 1 shows the code generation pipeline of our hybrid on/off-chain approach. From a set of textual constraints, regulations, or policies (e.g., health insurance), domain experts computerize *declarative N3 rules*. An *OWL2 domain ontology* encodes the semantics of concepts and relations found in well-known domain standards such as HL7 FHIR.

Given the OWL2 ontology, the **Parser** parses the N3 rules and extracts a set of annotated *rule graphs*. Next, the **Converter** converts these rule graphs into a set of intermediary *"bridge" abstractions* that represent the declarative logic using imperative programming concepts. Whereas N3 rules declare applicable data constraints, the imperative code procedurally implements and checks these constraints against input data. From these bridge abstractions, **Code Generators** will generate smart contracts for different blockchain languages (e.g., Solidity, JavaScript). Once the **Smart Contract** is deployed on a blockchain, it will be executed by a client using *transactions with input data* (e.g., insurance claim). During its execution, the contract will *request additional data* from distributed Information Systems (IS) (e.g., Electronic Medical Records) via an **off-chain oracle** [14]. Finally, the contract will issue *output events* that reflect the outcome (e.g., claim approved).

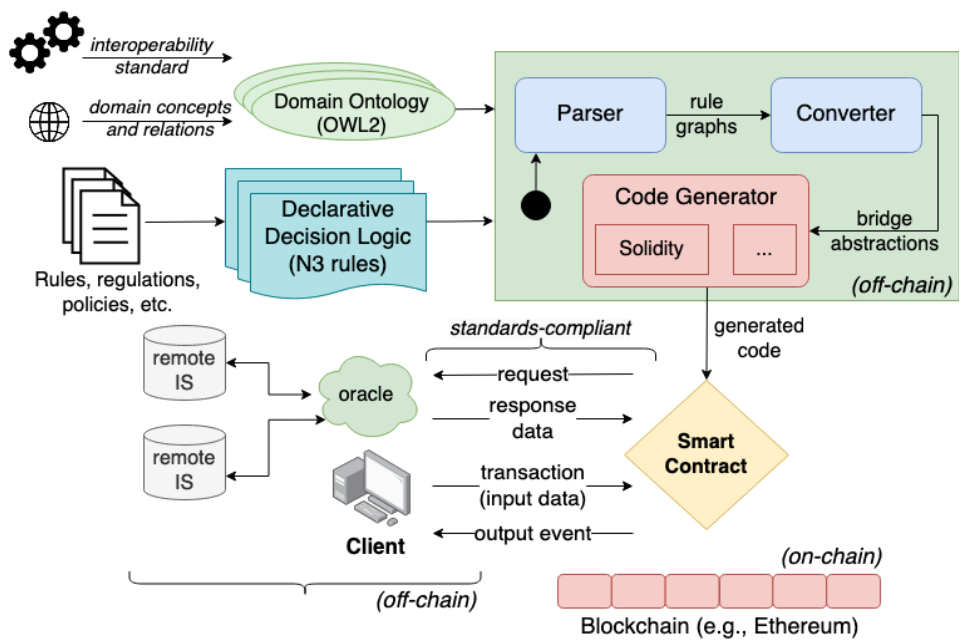

**Figure 1**. Architecture overview of the code generation approach.

Next, we show how high-level decision logic is encoded using N3 rules, and show the corresponding rule graphs. We introduce a set of bridge abstractions to represent these rule graphs in an imperative language. We discuss multiple considerations for code generators that output the final smart contract code.

# Representing Decision Logic using N3

We will rely on a health insurance use case as a running example. This use case involves deciding on reimbursement of healthcare costs, based on the patient's coverage and policies from the Medicare's booklet on benefits [20] (Section 1, p. 98). We show a text snippet of the case below; we will represent the numbered parts as N3 rules and use them to illustrate our code generation approach. We refer to our online repository for the full case [21].

"(1) Medicare covers transplant drug therapy if Medicare helped pay for your organ transplant. You must have Part A at the time of the covered transplant, and (2) you must have Part B at the time you get immunosuppressive drugs. Keep in mind, (3) Medicare drug coverage (Part D) covers immunosuppressive drugs if Part B doesn't cover them. [..]"

This text is interpreted as follows. (1) An organ transplant is eligible for drug therapy coverage in case Medicare Part A helped pay for the transplant. (2)-(3) Transplant drug therapy (immunosuppressive drugs) is covered in case you, at the time of therapy, have (2) Medicare Part B, or (3) if not Part B, you have Medicare Part D.

## Decision Logic As Existential N3 Rules

Notation3 (N3) [16] is a Semantic Web rule language that is built on top of the Resource Description Framework (RDF). N3 has been used in health informatics to represent clinical decision logic (e.g., extracted from guidelines) [22]–[24], explain the ensuing health recommendations [25], and generate model-driven UIs [26]. N3 supports a wide range of built-in operators, meta-data on groups of statements, and (scoped) negation as failure. There are execution engines for multiple platforms (eye [27], eye-js [28], cwm [29], jen3 [30]) and a VSCode plugin [31] for writing and executing N3 code.

In RDF [32], information is described as *triples* with subject, predicate, and object resources[4]. *URI* and *literal terms* (e.g., numbers and strings) can be used to identify resources; *blank nodes* are used when an identifier is not available or necessary. URIs can be written in full (e.g., `<http://hl7.org/fhir/Patient>`) or using qualified names written as `<namespace>:<localname>`. E.g., on line 1 in Code 1, the URI uses *fhir* as namespace and *CoverageEligibilityRequest* as local name; on line 3, the default (empty) namespace and *immunoTherapyItem* as local name is used. A triple is terminated by a period ("."). Shorthands include the predicate "a" for type (e.g., line 1); a single subject with multiple predicate-object pairs separated by ";" (e.g., lines 1 and 2); and a single subject-predicate pair with multiple objects separated by "," Blank nodes are written using "[" "]"[5].

N3 extends RDF with graph terms[6] for grouping triples, surrounded by "{" "}"; lists as first-class citizens, surrounded by "(" ")"; and universal variables, written as `?<name>`. An N3 rule is written as a triple: the subject is a graph term as the rule's body or antecedent, the predicate is *log:implies* (shorthand "=>"), and the object is a graph term as the rule's head or consequent. Simply put, given a rule { body } ⇒ { head }, statements in the head will be inferred in case statements in the body evaluate to true.

---

[4] Not to be confused with HL7 FHIR "resources", i.e., modular components to describe health data.
[5] Alternatively, labeled blank nodes can be used [32].
[6] We refer to the N3 CG report [16] for their formal meaning.

Code 1 is an N3 rule that checks whether a relevant coverage eligibility request was submitted. The rule uses concepts and relations from the HL7 FHIR ontology[7], a healthcare interoperability standard, and clinical domain ontologies (SNOMED-CT [33]):

```
1. { ?req a fhir:CoverageEligibilityRequest; fhir:purpose 'validation' .
2.   ?req fhir:item ?med . ?med a fhir:MedicationRequest ; fhir:medication "DBCAT005063"
3. } => { ?req :immunoTherapyItem ?med } ; cg:functionParam ?req .
```

**Code 1**. N3 rule supporting parts (2) and (3) of the health insurance example.

- Line 1: a coverage eligibility request (*?req*, type fhir:CoverageEligibilityRequest) has a purpose of validation;
- Line 2: the eligibility request *?req* has an item that is a medication request (*?med*, type fhir:MedicationRequest) for immunotherapy medication (Drugbank code DBCAT005063);
- Line 3: if so, the rule infers that *?req* has *?med* as an immuno-therapy item (line 3). We discuss the annotations (*cg:functionParam*, *cg:event*) in the rule graph section.

Code 2 encodes part (1) of our running example:

```
4. { ?req :immunoTherapyItem ?med ; fhir:patient ?p .
5.   { ?claim a fhir:Claim ; fhir:subject ?p }
6.         cg:request <http://myfhir.ca/> .
7.   ?claim fhir:procedure ?proc . ?proc fhir:status 'completed' ; fhir:code 77465005 .
8.   ?claim fhir:insurance ?cov . ?cov fhir:insurer :Medicare ; fhir:class :PartA .
9. } => { ?med :eligibleTransplant ?proc } ; cg:functionParam ?req .
```

**Code 2**. N3 rule encoding part (1) of the health insurance example.

- Line 4: a request (*?req*) has *:immunoTherapyItem ?med* (inferred from Code 1) and concerns patient *?p*;
- Lines 5-6: this graph term (surrounded by "{}") includes 2 triples that encode data requirements from a remote IS (`<http://myfhir.ca/>`), in particular, previous claims of patient *?p* (*?claim*, type fhir:Claim).
- Line 7: a previous *?claim* covered a procedure *?proc* with status completed and type transplant (SNOMED code 77465005);
- Line 8: the *?claim* had insurance coverage *?cov* that was insured by Medicare class Part A;
- Line 9: if all these conditions are met, the rule infers that the *?med* request item has a transplant *?proc* that is eligible for drug therapy coverage.

Code 3 encodes parts (2) and (3) of the running example:

```
10. { ?req fhir:patient ?p ; :immunoTherapyItem ?med . ?med :eligibleTransplant ?proc .
11.    { ?cov a fhir:Coverage; ?cov fhir:policyHolder ?p ; fhir:status 'active' } ;
12.         cg:request <http://myfhir.ca/> .
13.    ?cov fhir:insurer :Medicare ; fhir:class ?class . ?class list:in (:PartB :PartD).
14. } => { [] a fhir:CoverageEligibilityResponse ; fhir:request ?med ;
15.         fhir:outcome 'complete' } ; cg:functionParam ?req ; cg:event 'Response' .
```

**Code 3**. N3 rules encoding parts (2) and (3) of the health insurance example.

---

[7] For readability and without loss of generality, we show a less verbose version of FHIR. In particular, it does not rely as heavily on blank nodes and multi-valued properties.

- Line 10: a request *?req* concerns patient *?p*, has an immuno-therapy item *?med* (inferred from Code 1) and a transplant eligible for drug therapy coverage (inferred from Code 2);
- Lines 11-12: similar to before, this graph term includes 3 triples that encode data requirements from a remote IS (`<http://myfhir.ca/>`), in particular, active coverages *?cov* of patient *?p* (*?cov*, type fhir:Coverage);
- Line 13: a current coverage is insured by Medicare with class PartB or PartD[8];
- Line 14-15: if these conditions are met, the rule infers a successful coverage eligibility response that concerns the request item *?med*.

Code 3 is an existential rule as it includes a blank node in the rule head (indicated by "[ ]"). Each time the rule fires and the consequent is inferred[9], a new blank node will be created [34]; here, a new node of type fhir:CoverageEligibilityResponse with the given properties.

We point out that we encoded the conditions of the health insurance example using multiple N3 rules; each rule infers whether an individual condition was met. These rules then "chain" together to implement the decision logic, as subsequent rules refer to inferences of prior rules (e.g., "immunoTherapyItem" "eligibleTransplant"). This modularity allows for better code reuse; rules can easily be added for coverages other than those shown in Code 3, such as ESRD-related Medicare in the last 36 months (see full use case [21]) without having to duplicate Codes 1 and 2. Moreover, it allows keeping track of individual conditions: it is possible that a transplant was eligible for drug therapy coverage, yielding an "eligibleTransplant" inference, but the patient did not have Part B or Part D.

## Existential N3 Rules as Annotated Rule Graphs

The *Parser* component will parse the N3 rules from Codes 1-3 into a set of rule graphs. Figure 2 shows the rule graph corresponding to Code 3[10]. Our code generation pipeline leverages these graphs to generate intermediary bridge abstractions.

---

[8] The *list:in* builtin checks whether *?class* is found in a list, thus representing a (limited) disjunction.
[9] From a logical viewpoint, blank nodes are existentially quantified variables, making such rules existential rules. The consequent is inferred in case it is not yet present in the dataset.
[10] Rule graphs corresponding to Codes 1 and 2 are shown in Appendix A.

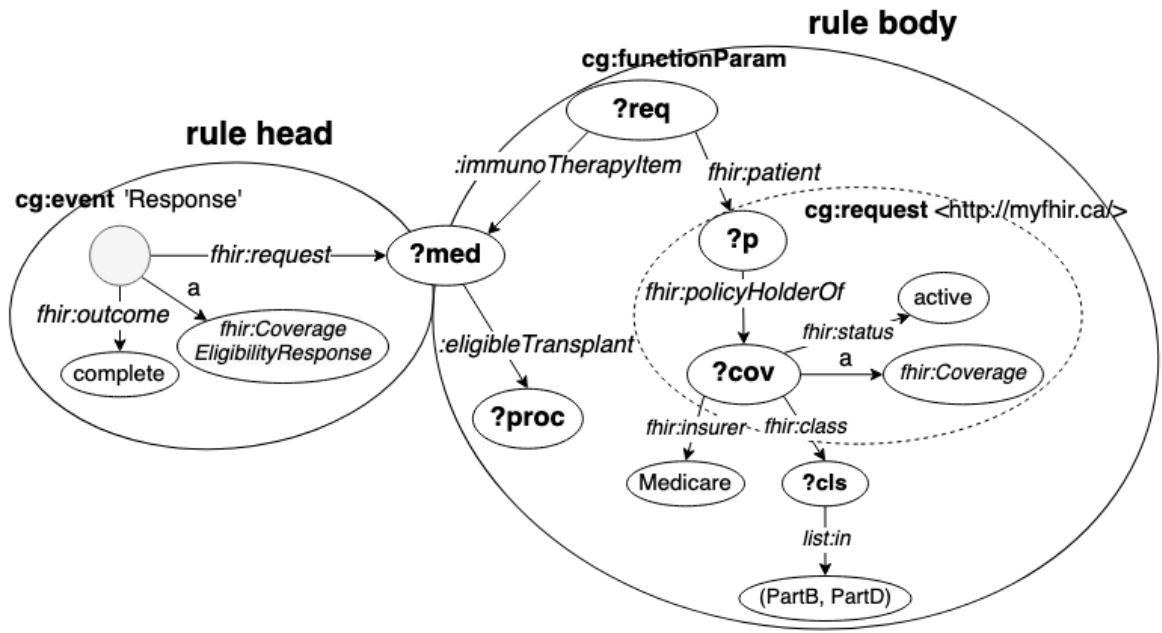

**Figure 2**. Rule Graph extracted from Code 3.

Edges in the rule graph are annotated based on the rule clause (head vs. body) they were found, together with any N3 rule annotations (bold text). The *cg:functionParam* annotation indicates the input parameter of the generated smart contract: in this case, the contract will process a given *?req*, i.e., a coverage eligibility request. The *cg:event* annotation indicates that, when all rule conditions are met, an event called "Response" will be emitted with the coverage eligibility response, with *fhir:outcome* complete and concerning the request item *?med*. The *cg:request* annotation indicates data to be retrieved during execution: coverages *?cov* of type *fhir:Coverage*, with *fhir:policyHolder ?p* (current patient) and *fhir:status 'active'*, will be retrieved from the remote IS with location `<http://myfhir.ca/>`.

Our approach currently places multiple restrictions on N3 rules:
(1) Regarding N3 rules, they should be structured as a "star-shaped" graph, originating from a function parameter and consisting of sequential node paths. For non-path structures, the *Parser* will attempt to identify inverse properties; e.g., in Figure 2, the inverse *fhir:policyHolderOf* property is used to link the parameter *?req* to coverages *?cov* using a sequential path. In case of multiple rules, they should form a single sequential chain that ultimately leads to desired inference(s).
(2) Regarding terms, all predicates should be concrete (non-variable), variables can occur as intermediary and leaf nodes, and concrete terms can occur as leaf nodes. Currently, a small number of N3 math operations (sum, product, quotient, exponent) are supported. Each URI term should further correspond to exactly one ontology class declaration (e.g., type *Patient* for node *?p*), as this simplifies their correlation with ADTs (Algorithm 1).

## Converting Rules into Imperative Bridge Abstractions

The semantic KG—including the extracted rule graphs and domain ontology—is converted into a set of intermediary "bridge" abstractions, which represent the declarative logic using imperative programming concepts. In the declarative interpretation, a rule of the form *condition→consequence* will describe that meeting the *condition* implies the truth of the *consequence*; e.g., when a patient is over 65, then they are elderly. The imperative application logic will reflect an operational interpretation; e.g., when a patient is over 65, assigning property "elderly" to the patient.

The *Converter* component starts by generating *a minimal set of ADTs* (Figure 3) to structure incoming data and outgoing events. Subsequently, the *Converter* generates imperative application abstractions (Figures 3-5) to operate on these ADTs to process the data and issue events. At a later stage, these ADTs and abstractions will be converted into a concrete blockchain smart contract. The full smart contract code generated for the N3 rules can be found in our online repository [21].

## Constructing ADT from Rule Graphs and Domain Ontologies

Figure 3 shows the set of ADT bridge abstractions as typically found in imperative applications[11].

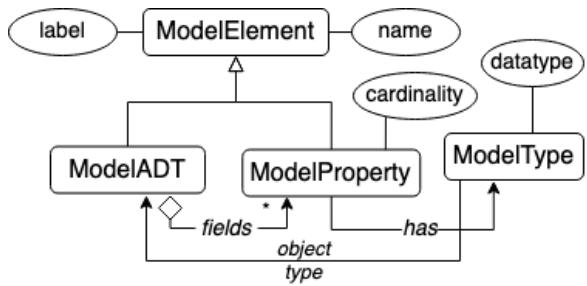

**Figure 3**. Bridge Abstractions: Abstract Data Types.

A `ModelADT` instance represents a concrete ADT (e.g., *Patient*) that corresponds to an ontology class referred by the rule. It consists of `ModelProperty` fields (e.g., *status*, *gender*) with a given cardinality, corresponding to the ontology class's properties referred in the rule. To support statically typed languages (e.g., Solidity), a `ModelProperty` keeps a `ModelType` representing either an *XSD datatype*[12] or an *object type* as another `ModelADT`. `ModelADTs` can thus be nested, i.e., the value of a `ModelProperty` can be another `ModelADT`. Each `ModelElement` keeps its term URI from the rule or ontology as a unique name and a human-readable `label` (*rdfs:label* property in the ontology), if any.

To instantiate these ADTs, our approach recursively traverses the extracted rule graph(s) (e.g., Figure 2) starting from their function parameter (e.g., *req*), instantiating ADTs for classes and properties referred by the rule. This is in contrast to other work such as RDFReactor [35], which instantiates ADTs (Java classes) for an entire OWL ontology. This aims to reduce the size of smart contracts. For instance, on the Ethereum blockchain, smart contracts have max. size of 24Kb[13], whereas mature ontologies can be huge; DMTO [36], used in our prior work to generate diabetes-related smart contracts [8], includes over 10,000 classes. Generating ADTs for all ontology classes would yield a smart contract far exceeding the max. size.

---

[11] Note that these are not sufficiently general to cover all possible imperative code; rather, they cover the code needed to implement decision logic from restricted N3 rules.
[12] See https://www.w3.org/TR/xmlschema-2.
[13] Note that each byte of storage on blockchain expends cryptocurrency as well.

Algorithm 1 shows the simplified pseudocode of the *generateADT* function. This function is called for the function parameter (*node* argument) and refers to the abstractions from Figure 3. The end result of this function for two ADTs is illustrated in Code 5. For each graph node, its ontology type is established (*getOntologyType*), based on its type from the rule or property domains/ranges from the ontology (line 3). E.g., in Code 3, *?p* has class *Patient* as defined by the ontology (range of the *patient* predicate); *?claim* has class *Claim* as found in the rule. In case of an ontology class (lines 6-11), either its prior `ModelADT` will be reused (lines 7-8), or a new `ModelADT` will be instantiated (lines 10-11). We discuss lines 4-5 below.

For the node's outgoing edges that represent class properties, i.e., not comparators or *rdf:type* properties, a new `ModelProperty` is added to the current `ModelADT` (lines 12-14). E.g., for ADT *Coverage*, properties *status*, *policyHolder*, *insurer,* and *class* are added. A `ModelProperty` will have a cardinality based on the OWL ontology (cardinality constraints [17] and functional property annotations[14]). Its `ModelType` will be based on the target node, which is established by recursively calling *generateADT*. In case the target node is a datatype, the function returns the corresponding `ModelType` (lines 4-5); otherwise, the function will return a `ModelType` with the current ADT (line 15). For instance, *Coverage*'s *policyHolder* property has type *Patient*, and *status* has datatype *string*.

```
1.  global: ruleModel: [ModelADT]
2.  function generateADT(node):
3.     type ← getOntologyType(node)
4.     if type is datatype then # datatype
5.        return ModelType(type)
6.  else # ontology class
7.     if ruleModel[type] then
8.        ADT ← ruleModel[type]
9.     else
10.       ADT ← ModelADT(type, node)
11.       ruleModel ∪ ← ADT
12. for edge in node.outgoing that is not comparator or rdf:type do
13.    edgeType ← generateADT(edge.target)
14.    ADT.fields ∪ ← ModelProperty(edge.term, edgeType)
15. return ModelType(ADT)
16. model ∪ ← ruleModel
```
**Algorithm 1**. Generate Abstract Data Types (ADT) Pseudocode.

```
ADT CoverageEligibilityRequest:
  item [0..*]: MedicationRequest
  purpose [0..1]: string
  immunoTherapyItem [0..1]: MedicationRequest
  eligibleTransplant [0..1]: Procedure

ADT Coverage:
```

---

[14] For now, we assume a unique-name assumption (meaning 2 different URIs always refer to different RDF resources), as this simplifies the interpretation of cardinality constraints.

```
    policyHolder [0..1]: Patient
    status [0..1]: string
    insurer [0..1]: Insurer
    class [0..1]: Class
```
**Code 5**. ADT Pseudocode for CoverageEligibilityRequest, Coverage

After processing a rule, the generated ADTs (`ruleModel`) are merged into the overall ADT model (line 16). E.g., the `CoverageEligibilityRequest` ADT includes properties from rule 1 (`item`, `purpose`, `immunoTherapyItem`) and rule 2 (`eligibleTransplant`).

## Constructing Application Logic from Rule Graphs

Figure 4 shows a basic set of bridge abstractions for application logic as typically found in imperative applications[15].

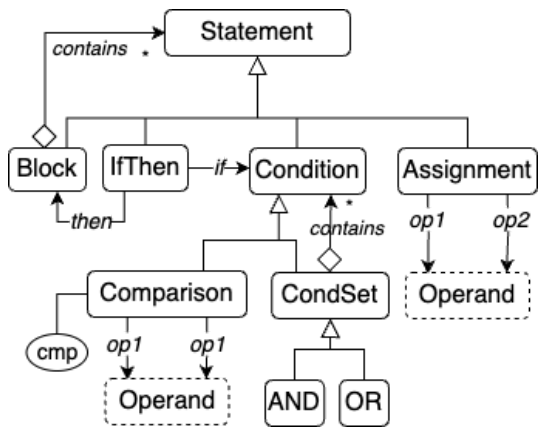

**Figure 4**. Bridge Abstractions: Application Logic - Basic Statements.
(Dashed lines indicate that the element is elaborated elsewhere.)

The generated application logic consists of a set of `Statements`. An `IfThen` statement keeps an "if" `Condition` and a "then" `Block`. A `Block` acts as a container of other `Statements`. A `Condition` can be a `Comparison`, which compares two `Operands` (e.g., literals; see below) using a comparator (`cmp`); or a `CondSet`, which groups multiple `Conditions` as either a conjunction (`AND`) or disjunction (`OR`)[16]. An `Assignment` assigns an operand to another operand (e.g., variable). Since a `Block` is itself a `Statement`, other `Blocks` (or any type of statement) can be arbitrarily nested within `Blocks`; similarly, as `CondSet` is a `Condition`, conditions can be arbitrarily nested.

Figure 5 shows the statement operand abstractions:

---

[15] As before, we note that these are only meant to cover the code needed to implement decision logic from restricted N3 rules.
[16] As mentioned, a limited type of disjunction is supported based on the *list:in* builtin.

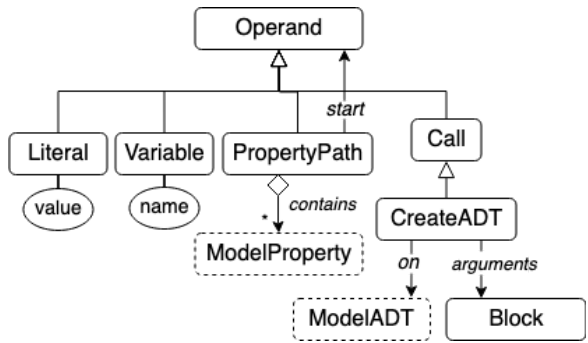

**Figure 5**. Bridge Abstractions: Application Logic - Operands.

An `Operand` can be a `Literal` (e.g., string or number), `Variable`, `PropertyPath`, or an ADT constructor (`CreateADT`) as a type of `Call`. A `PropertyPath` starts from an operand (typically a variable) and consists of a sequence of `ModelProperties` to navigate through nested `ModelADTs`, with the goal of identifying a specific (set of) value(s) (see Code 6 for an example). A `CreateADT` instance represents the invocation of a `ModelADT` constructor, with a set of arguments given as a `Block`.

Similar to before, to instantiate these imperative application concepts, our approach recursively traverses the extracted rule graph(s) starting from their function parameter. In a nutshell, every rule will be converted into an `IfThen` instance: the rule body is converted to the IF part (`Conditions`) and the rule head into the THEN part (`Statements`). Considering Figure 3, for each node path shown, a `PropertyPath` will be created, such as `req.patient.policyHolderOf.insurer`; if found in the rule body (IF part), the path will be used in a `Comparison`; otherwise (THEN part), an `Assignment`. The end result for Figure 3 is shown in Code 6. Extensions to this function for remote data requests, multi-valued properties, operations, and unification are discussed in the following subsections.

Algorithm 2 shows the simplified pseudocode of the *generateLogic* function, which is called for each function parameter (e.g., *?req*). The *node* argument represents the current graph node (e.g., *?req*); *from* is the graph edge *leading to* this node, and the *path* is the current `PropertyPath`. This *path* is constructed as the *generateLogic* function is called recursively for outgoing edges (lines 21-24). The collected *conditions* (Conjunction; IF part) and *block* (Block; THEN part) are passed as well.

Given a graph node, we retrieve the corresponding `ADT` as created by the *generateADT* function (line 4). In case the node term is concrete (i.e., not a variable), we construct a corresponding `Literal` (lines 5-11). If found in the rule body, we create a `Comparison` between the current *path* and the literal (lines 7-9); it is possible a comparator was specified in the rule (incl. *list:in*; Code 3), otherwise, an equality comparator is used (line 7). If found in the rule head, an `Assignment` is created (lines 10-11). For blank nodes in the rule head (lines 13-18), we point out that a new blank node will be created whenever the rule fires. To represent this behavior, we instantiate a new constructor invocation (`CreateADT`) of the node's `ModelADT`. We then assign this new ADT to the current path. Any paths starting from this node will serve as arguments to the constructor; e.g., in Figure 2, the *fhir:outcome* path from the blank node will act as a constructor argument (see Code 6, line 7). To that end, the current path is reset (line 17) and subsequent *generateLogic* calls will use the constructor's

*arguments* block (line 18), i.e., assignments will be added as arguments. In case the rule is annotated with *cg:event*, an event emitting the new ADT is inserted as well (not shown for brevity). Finally, for any variable in the rule body (lines 19-20), we add an extra Comparison that ensures an ADT exists at the end of this path. As it is possible that a property does not have any value, this avoids null pointer exceptions when the path is traversed at runtime.

This way, Code 3 is converted into the application code shown in Code 6. We discuss remote data requests and unification (required by line 7) in the following subsections.

```
1. global: model:[ModelADT], conditions:Conjunction, block:Block
2. function generateLogic(node:GraphNode, from:GraphEdge, path:PropertyPath,
3.              conditions:Conjunction, block:Block):
4.    ADT ← model[node]
5.    if node.term is concrete then
6.       concrete ← Literal(node.term)
7.       if from.source is 'body' then
8.          cmp ← if from.term is 'comparator' then from.term else 'equal'
9.          conditions ∪ ← Comparison(path, cmp, concrete)
10.      else # rule head
11.         block ∪ ← Assignment(path, concrete)
12.   else # variable
13.      if from.source is 'head' then
14.         if node.term is blank node then
15.            constructor ← CreateADT(ADT)
16.            block ∪ ← Assignment(path, constructor)
17.            path ← PropertyPath()
18.            block ← constructor.arguments
19.         else # rule body
20.            conditions ∪ ← Comparison(path, 'exists')
21.   for edge in node.outgoing do
22.      if edge.term is not 'comparator' and is not rdf:type then
23.         path2 ← path ∪ ADT.fields[edge.term]
24.         generateLogic(edge.target, edge, path2, conditions, block)
25. logic ∪ ← IfThen(conditions, block)
```
**Algorithm 2**. Generate Application Logic Pseudocode.

```
1. if req.immunoTherapyItem exists
2.     and req.immunoTherapyItem.eligibleTransplant exists
3.     and req.patient exists and req.patient.policyHolderOf exists
3.     and req.patient.policyHolderOf.insurer == 'Medicare'
4.     and req.patient.policyHolderOf.class in ('PartB', 'PartD')
5. then
6.     v = CoverageEligibilityResponse(
7.            request: req.immunoTherapyItem, outcome: 'complete')
```
**Code 6**. Application logic pseudocode for Code 3

Afterwards, per rule, an IfThen instance is created with the generated *conditions* and *block* (line 22). The application *logic* will be inserted into a single function, which accepts as argument(s) the annotated function parameter(s) (e.g., req).

## Supporting Advanced Application Logic

Figure 5 shows the statements that cover more "advanced" application logic:

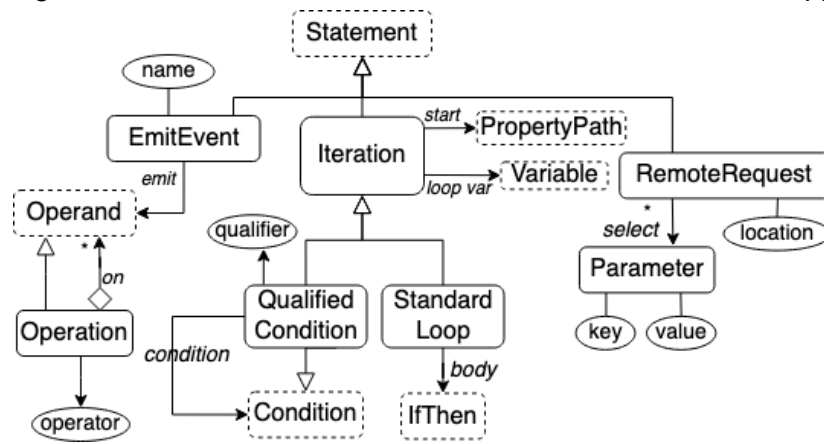

**Figure 6**. Bridge Abstractions: Application Logic - Advanced Statements.

An `Operation` (e.g., math or string operation) is applied to a set of `Operands` using an `operator` (e.g., sum). An `Iteration` will start from a `PropertyPath`: in each iteration, a value identified by the path (`start`) is assigned to the loop's `Variable`, and a `StandardLoop` will execute its body (`IfThen`). A `QualifiedCondition` will check the given `Condition` and will return true if the condition holds for all (universal qualifier) or some (existential qualifier) loop variable values. A `RemoteRequest` represents a data retrieval request for a remote IS at `location` with key-value `Parameters` for selecting data. Finally, an `EmitEvent` represents the emission of an `Operand` as an event with a given `name`. We point out that `Operation` is itself an `Operand`, meaning it can be nested in comparisons or other operations; and a `QualifiedCondition` is a `Condition`, meaning it can be used wherever conditions are used.

In the subsections below, we discuss extensions to the *generateLogic* function to support the instantiation of these abstractions as well as multi-valued properties.

### Remote Data Requests

In our distributed Health 3.0 setting, relevant data will need to be retrieved from remote IS to implement the decision logic, such as prior claims and coverages of the given patient.

In general, for smart contracts on blockchain to communicate with remote services, an off-chain "oracle" is needed [14]. Communication between the smart contract, oracle, and remote service takes place as follows:

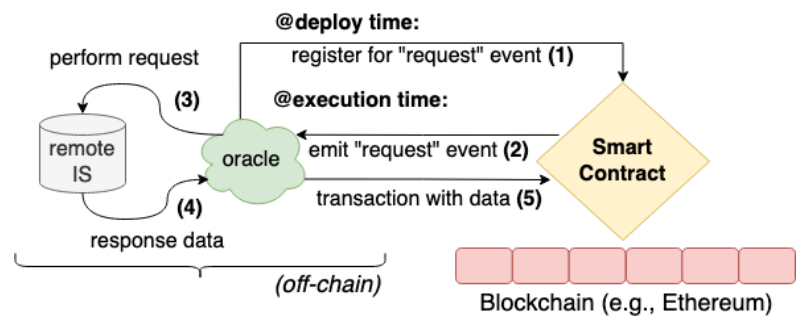

**Figure 7**. Oracles for off-chain communication.

After the smart contract is deployed, the oracle registers for the contract's "request" event (1). At execution time, the contract will emit this event when remote data needs to be retrieved (2); the event is received by the oracle, which performs the request (3) and receives the data (4) from the remote IS; subsequently, the oracle issues a transaction with the received data to the smart contract (5).

To represent remote data requests, the N3 rules include graph terms annotated with *cg:request*. We show the graph term from Code 3 below:

```
{ ?cov a fhir:Coverage . ?cov fhir:policyHolder ?p ; fhir:status 'active' . };
    cg:request <http://myfhir.ca/> .
```

In general, the graph term annotated with *cg:request* is structured as follows:

```
1. {   ?x a <type> .
2.     ?x <property 1> <value 1> ... ?x <property n> <value n> };
3.     cg:request <location> .
```

Where:

- *?x* is the "remote entity" about which data is retrieved, e.g., *?cov*.
- *<type>* identifies the type of data requested, e.g., *fhir:Coverage*.
- *<property>* / *<value>* pairs select subsets of the remote data, where *<value>* can be a concrete value or a variable occurring elsewhere in the rule. E.g., property-value pairs *fhir:policyHolder* / *?p* and *fhir:status* / *active* select a set of coverages with the current patient as policyholder and an active status.
- The *cg:request* value indicates the location of remote IS from which the data will be retrieved, e.g., `<http://myfhir.ca/>`.

Our health insurance use case utilizes HL7 FHIR as an interoperability standard, which offers a RESTful API for retrieving FHIR-compliant data[17]. Based on the above, a suitable GET URL can be constructed as follows:

```
<location>/<type>/_search?<property 1>=<value 1>&...&<property n>=<value n>
```

Appendix B shows the pseudocode that extends the *generateLogic* function with remote data requests. Plugging in a remote data request leads to the following for Code 3:

```
1. if req.immunoTherapyItem exists
2.     and req.immunoTherapyItem.eligibleTransplant exists
3.     and req.patient exists then
4.   r = RemoteRequest("myfhir.ca", { policyHolder=req.patient, status='active' })
5.   if r.insurer == 'Medicare'
6.       and r.class in ('PartB', 'PartD')
7.   then
8.     v = CoverageEligibilityResponse(
9.           request: req.immunoTherapyItem, outcome: 'complete')
```

**Code 7**. Application logic pseudocode for Code 3 with remote service invocation.

Note that, in case multiple results are returned, they will have to be iterated over; we elaborate on iteration next.

**Multi-Valued Properties Using Indexing and Iteration**

---

[17] https://build.fhir.org/http.html#search

Until now, we have made the simplifying assumption that properties have either none or a single value. However, properties can have any cardinality (Figure 3); we call properties with multiple possible values "multi-valued". E.g., in the FHIR standard, a *CoverageEligibilityRequest* can include many items (*fhir:item*); a *MedicationRequest* can cover multiple medications (*fhir:medication*); and a patient can have many prior insurance claims (inverse *fhir:subjectOf* property).

As with our choice for a hybrid on-/off-chain pipeline, our implementation is informed by the economic rules of blockchain-based systems, i.e., where computational work expends cryptocurrency. Where possible, values of multi-valued properties are indexed to avoid iterations (time complexity $O(n)$) over the values. We discuss two general methods for processing multi-valued properties. Appendix C elaborates on their implementation. We discuss the challenges of implementing these methods in Solidity in the next section.

*Indexing Multi-Valued Property Values on Unique Keys*

In case the decision logic refers to values of multi-valued properties using a unique key, we can index and access these values using those keys. Below is an example snippet from our prior work [8], where `DDO:has_physical_examination` is a multi-valued property:

```
1. { ?profile DDO:has_physical_examination ?exam .
2.     ?exam rdf:type DDO:BMI ;
3.         DDO:has_qualitative_value ?value .
4.     ?value math:notLessThan 23 }
```
**Code 8**. Example multi-valued property that allows indexing on a unique key.

The rule refers to a value of `DDO:has_physical_examination` using its unique key, i.e., the type of examination[18] (BMI), and then compares it to a constant value. Here, we can (a) index multi-valued property values on their key using a dictionary and then (b) generate application logic that accesses unique values from this dictionary:

```
  profile.has_physical_examination['BMI'].has_qualitative_value >= 23
```
**Code 9**. Application logic pseudocode for Code 8.

*Iterating over Multi-Valued Property Values*

In case a unique key is not available for values of multi-valued properties, we will have to iterate over its values. From our running example, *fhir:item* is an multi-valued property:

```
1. { [..]
2.     ?req fhir:item ?med ; ?med a fhir:MedicationRequest ; fhir:medication "DBCAT005063"
3. } => { ?req :immunoTherapyItem ?med } ; cg:functionParam ?req .
```
**Code 10**. Example multi-valued property that requires iterating over its values.

The rule cannot refer to a value of *fhir:item* using a unique key; the type of item (e.g., *MedicationRequest*) is not unique, as there can be multiple items of the same type. As an alternative, a randomly generated id could be used, but this would not accommodate the decision logic; the rule is interested in all medication requests, not items with a specific id. In this case, we (a) keep the values using an array and then (b) generate application logic to iterate over the array:

```
1.  for each req.item as item do
2.    if item.type == 'MedicationRequest'
3.      and item.medication.concept == "DBCAT005063" then
4.        req.immunoTherapyItem = item
```
**Code 11**. Application logic pseudocode for Code 10

---

[18] In this case, only a single examination per type was kept, making the type value a unique key.

For each value (*item*) of path *req.item* (line 1), if the *item* has type *MedicationRequest* (line 2) and the medication's terminology concept equals DBCAT005063 (immunotherapy) (line 3), then we assign the item to the path *req.immunoTherapyItem*.

If the *fhir:medication* property is assumed to be multi-valued, we further require an existential qualified condition in line 3:

```
1.  for each item in req.item do
2.    if item.type == 'MedicationRequest' and
3.      for some med in item.medication holds that med.concept == "DBCAT005063"
4.    then
5.      req.immunoTherapyItem = item
```
**Code 12**. Application logic pseudocode for Code 10 (2)

This existential qualified condition checks whether any of the *item*'s medications (*item.medication*; *med*) has a concept indicating immunotherapy; if so, then the aforementioned assignment takes place.

### Unification and Mathematical Operations

During code generation (Algorithm 2), path sequences from rule graphs are converted into property paths for comparison or assignment with concrete values or new ADTs. In some cases, however, *property paths will have to be compared or assigned to each other*.

In Code 2 (line 9), the `req.immunoTherapyItem` property path identifies values for the `?med` variable, and `req.patient.subjectOf.procedure` identifies values for `?proc`. To represent this in imperative code, the following assignment should be generated:

`req.immunoTherapyItem.eligibleTransplant = req.patient.subjectOf.procedure`

Similarly, in Code 3 (line 14), the `?med` variable is identified by the `req.immunoTherapyItem` property path, and should lead to the following constructor invocation:

`CoverageEligibilityResponse(request: req.immunoTherapyItem [..])`

We refer to resolving variables `?proc` and `?med` to their corresponding property paths as unification. We implement the unification step by keeping a mapping between variables, found in the rule body, to their property path. When the same variable is found in the rule head, it is resolved to the mapped path. We expand the *generateLogic* function as follows[19], assuming a *var_map* global:

```
19       else # variable in rule body
u1.        var_map[node.term] ← path # map the variable to its property path
```

And

```
13.      if from.source is 'head' then # variable in rule head
14.        if node.term is blank node then
             ...
u2.        else # (non-blank) node variable in rule head
u3.          path2 ← var_map[node.term] # get property path mapped to variable
             # create assignment
u4.          assignments ∪ ← Assignment(path, path2)
```

**Algorithm 4**. Unification added to the generateLogic function.

We use the same unification mechanism to support mathematical and other operations and their nesting. This is elaborated in Appendix D.

## Generating Imperative Smart Contract Code

The *Code Generation* component will convert the instantiated bridge abstractions into blockchain smart contracts. Currently, we support Solidity and JavaScript as target imperative languages. For brevity, we do not detail this step in this paper, but we describe noteworthy aspects below.

### Emit smart contract events

To communicate with any off-chain party, albeit the client that invoked the contract, or an oracle retrieving remote data, the smart contract emits events.

For communicating results back to the client, N3 rules are annotated with *cg:event <name>*, which will involve the emission of an event called *<name>*. An example is given in Code 3 (line 15). In case the application logic involved the creation of a new ADT, as is the case here, the event will emit this new ADT as payload:

```
emit Response(
    CoverageEligibilityResponse({ outcome: "complete", request: req.immunoTherapyItem })
);
```

### Remote data requests using oracles

Smart contracts follow a communication setup for remote data retrieval that relies on oracles, shown in Figure 7. The smart contract will emit an event when remote data is required; the oracle retrieves the data, and issues a transaction back to the contract with the data. In particular, this transaction calls a contract function, passing the retrieved data as an argument. Hence, the code relying on the retrieved data needs to be split into a separate function, which will then be called by the oracle. From our running example:

```
function process(CoverageEligibilityRequest memory req) public { ...
  emit OracleRequest(
    RequestData(req: req, {resource:"Claim", parameters:... }, callback:"callback1"));
}
function callback1(CoverageEligibilityRequest memory req, Claim[] memory data) public {
  for (uint i = 0; i < data.length; i++) {
    Claim memory v = data[i]; ...
}
```

---
[19] This pseudocode is simplified, i.e., only considering cases where variables in the rule head are unified with property paths from the rule body.

The original *process* function emits a data request to the oracle (*OracleRequest*), passing the original coverage eligibility request (*req*), the data request (type of HL7 FHIR resource, parameters), and the function to be called by the oracle (*callback*). This callback function includes code that relies on the remote data; it will be called by the oracle with the original request (*req*) and result data (*data*), and then processing will continue.

## Multi-Valued Properties using Indexing and Iteration

### Indexing Multi-Valued Property Values on Unique Keys

A unique challenge of blockchain-based systems is that computational work costs "gas" (i.e., expends cryptocurrency); as a result, blockchain languages tend to impose unique restrictions. In Solidity, loops are highly discouraged as each iteration may consume a different amount of gas, and unbounded loops may easily run into "out-of-gas" exceptions.

Hence, to implement multi-valued properties, we opted to use mapping and indexing where possible. However, mapping constructs are themselves restricted in Solidity: an ADT ("struct") that keeps a mapping cannot be nested within another struct. An example can be found in our prior work [8]:

```
struct Patient { PatientProfile profile; ... }
struct PatientProfile {
    mapping(DemographicConstants => Demographic) hasDemographic; ... }
```

To circumvent this issue, starting from the "root" struct (Patient in this case), we recursively merge the properties of all nested ADTs keeping mappings (e.g., PatientProfile) into the "root" struct. We update the property paths within the application logic accordingly.

### Iterating over Multi-Valued Property Values

Standard loops over arrays of values (`StandardLoop`, Figure 6) can be implemented using a regular for-loop in Solidity. However, to implement qualified conditions, albeit universal or existential (`QualifiedCondition`), Solidity does not have a built-in way to check a given condition for all or some elements of an array[20]. Hence, we need to define a separate function that implements the qualified condition, returning true or false based on the outcome. For the existential qualified condition in Code 12 (line 3), the following function would be generated[21]:

---

[20] Modern languages tend to offer functions (e.g., "every" and "some") to implement these directly.
[21] Comparing strings in Solidity also requires additional work that we leave out here for brevity.

```
function check1(Medication[] memory medications) private pure returns (bool) {
    for (uint i = 0; i < medications.length; i++) {
        Medication memory v1 = medications[i];
        if (v1.concept == "DBCAT005063")) { return true; }
    }
    return false; }
```

The function iterates over the array values one by one: if at least one value adheres to the associated condition (i.e., its concept equals a given string), true is returned, otherwise false. This function will then be called in the relevant "if" clause.

# Results

We used our code generation pipeline to generate smart contracts for 3 health insurance scenarios, including our running example. These smart contracts automate the process of deciding on reimbursement of healthcare costs, based on the Medicare's booklet on benefits [20]. We evaluated the generated smart contract code in 2 ways:

1) *code analysis*: visualizing their control and message flow, and using simulated cases to validate their correctness.

2) *performance analysis*: measuring their execution cost in terms of "gas", i.e., computational effort required to execute transactions or smart contracts on a blockchain platform.

We refer to our online repository [21] for descriptions of these cases, their N3 implementation, generated smart contracts, and all evaluation results. We further deployed each contract on an Ethereum testnet called Sepolia so they can be manually tested. Experimental reproducibility criteria are available in our repository's README.

## Experiment Setup

We evaluated the generated smart contracts in (1) a local simulation environment, using Ganache [37] and Foundry [38] (analogous to a private consortium setting) and a (2) public test network, similar to the evaluation by [9]. We used web3.js [39] for writing client and oracle scripts that interact with our smart contracts. We describe these two scripts below:

**Oracle**. The oracle script is an off-chain component that listens to events that represent remote data requests from the smart contract. When received, the oracle retrieves the requested data, and returns the data to the smart contract by invoking its callback functions. In a real-world application, the oracle would fetch actual data from external sources (e.g., using HL7 FHIR RESTful APIs) instead of using hardcoded values as simulated here. However, we argue that this is a good approximation of how smart contracts can interact with external systems to obtain remote, off-chain data.

**Client**. The client script simulates an off-chain component interacting with the smart contract. It sends a request to the contract, and listens for an event that represents the contract's response, i.e., the outcome of the contract's decision logic. It thus acts as an interface between the user—or an external system—and the smart contract.

# Smart Contract Code Analysis

## Visualizing Control and Message Flow

The generated smart contracts will be part of a distributed Health 3.0 data environment, involving other contracts and off-chain components such as external data sources (oracles) and clients. Smart contracts use events to communicate with these off-chain components, resulting in an asynchronous workflow. For our running example, Figure 8 (a) shows a sequence diagram of the messages exchanged between *Client*, *Smart Contract* and *Oracle*. Figure 8 (b) shows actual internal and external function calls[22] recorded by Surya [40], a tool for visualizing smart contract control flow and highlighting vulnerabilities.

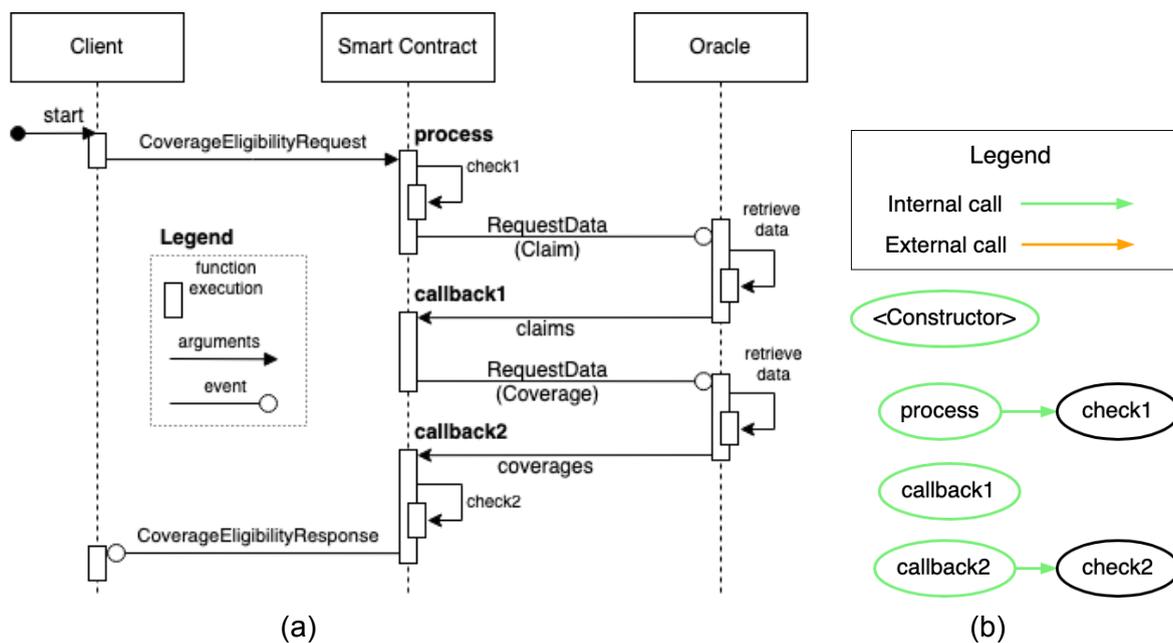

(a) (b)

**Figure 8**. (a) Sequence diagram of *Client*, *Smart Contract* and *Oracle*. (b) Function calls recorded by Surya[23] [40].

The main control flow proceeds as follows. The Client starts by invoking the *process* function of the Smart Contract, passing a *CoverageEligibilityRequest*. The process function then executes a number of if-else checks, including an internal *check1* function to check a condition on a multi-valued property. The process function proceeds by emitting a *RequestData* event to retrieve prior Claims. The Oracle listens for this event, retrieves the relevant claims, and calls the *callback1* function with the retrieved claims. This function then continues with processing these claims, eventually emitting a second *RequestData* event for prior Coverage. As before, the Oracle similarly responds by retrieving relevant coverages and passing them to a *callback2* function. The function processes these coverages, calls a second internal *check2* function, and finally emits a *CoverageEligibilityResponse* which is captured by the Client. In Appendix E, we describe all the structs, enums, events and functions from the smart contracts, including the ones mentioned here.

---

[22] Surya only captures function calls initiated by the smart contract, i.e., no event emissions or external calls of smart contract functions.
[23] Generated diagram re-drawn for legibility. We refer to Appendix E for the original diagrams for all 3 smart contracts generated by Surya.

## Validating Code Correctness

We created 10 test cases for each smart contract: for each case, we created CoverageEligibilityRequests (Client), Claims and Coverages (Oracle), together with their nested structs such as Procedure, Claim, MedicationRequest, and Patient (Appendix E). For each case, using the client and oracle scripts, we executed the smart contracts in our local simulation environment, and manually verified the correctness of the outcomes. All outcomes and their input data can be found in our online repository [21]. The generated contracts can also be tried out on the Sepolia testnet.

## Gas Analysis

Table 1 shows the execution costs of deploying and running the 3 smart contracts on a local simulation environment (Foundry [38]) in terms of units of "gas." Foundy's "gas- report" outlines the gas used by each function call and transaction. Deployment cost is the amount of gas used to deploy the contract onto the Ethereum blockchain.

| Contract 1* | | Contract 2 | | Contract 3 | |
|---|---|---|---|---|---|
| *deployment* | 1 234 470 | deployment | 1 584 046 | *deployment* | 1 453 908 |
| *process function* | 64 206 | *process function* | 157 363 | *process function* | 158 406 |
| *callback1 function* | 20 363 | *callback1 function* | 55 542 | *callback1 function* | 55 317 |
| *callback2 function* | 20 312 | | | | |

**Table 1**. Deployment and execution costs as units of "gas" in a local network.
* This is the smart contract for our running example.

On Ethereum blockchains, each computation costs "gas," which is paid in Gwei. 1 Gwei is one-billionth of ETH (the Ethereum cryptocurrency). The gas costs for atomic operations (e.g., base transaction fees, creating a smart contract, calling functions) are detailed in Appendix G of the Ethereum Yellow Paper [41]. E.g., the base cost of deploying a contract amounts to 53,000 gas, whereas storing the contract costs 20,000 gas per 256 bits of code.

Between January 2024 and the time of writing (Feb 22, 2024), the average gas price was circa 31 Gwei[24]. Deploying the first contract thus expends 38 268 570 Gwei; with 1 Gwei equalling one-billionth of ETH, this equals 0.03826857 ETH. Also at the time of writing (Feb 22, 2024), 1 ETH equaled 2 929.39 USD, implying a deployment cost of circa 112 USD. Calling the process, callback1 and callback2 functions incurs 0.001990386 ETH (5.83 USD), 0.000631253 ETH (1.85 USD) and 0.000629672 ETH (1.84 USD), respectively. Although not trivial, we argue that these costs are acceptable; deploying a reasonably-sized smart contract of 8Kb would already cost circa USD 157. Harris [42] measured the gas cost of deploying AI models on blockchain; when plugging in our Gwei cost per unit of gas and USD for ETH rate, deploying the cheapest model (Sparse Perceptron) amounts to USD 2812.

---
[24] https://ycharts.com/indicators/ethereum_average_gas_price (accessed Feb 22, 2024)

Public networks can suffer from variable and higher gas costs during periods of congestion. This could lead to unpredictable operating costs for health insurance applications. We point out that there are different types of blockchains: in a public blockchain, anyone can join and perform transactions. An alternative is a consortium chain: this chain is managed by multiple organizations and requires permission to join, and is typically used by banks and government organizations [43]. Importantly, a private consortium chain can be optimized for higher transaction throughput at lower costs. We discuss these types of blockchains, and their suitability for our health insurance use case, in the Discussion section.

# Discussion

## Healthcare Use Cases Benefiting from Blockchain

According to a recent literature study [44], research on the use of blockchain in healthcare has been growing since 2016. Automating health-related decision making using blockchain lowers administrative costs and avoids mistakes and delays caused by human error. At the same time, however, the choice of blockchain technology introduces latencies unrelated to computational work: these can be caused by the unavailability of nodes, chosen consensus mechanism (e.g., proof-of-work or -stake), and other blockchain properties, such as the use of interactive zero-knowledge proofs in preserving privacy. Moreover, any computational work incurs an execution cost on public blockchains in terms of cryptocurrency, as shown in our Results section.

A use case should thus have a distinct need for the unique "trustless" features offered by blockchain that allow it to act as a neutral intermediary (e.g., transparency, tamper-proofness). The National Institute of Standards and Technology (NIST) published a flowchart that determines the suitability of blockchain for a given use case [45] (p. 42). Criteria include (1) trust issues over who controls the dataset and runs the application logic, (2) a shared and consistent dataset between multiple entities; and (3) an immutable and tamper-proof transaction ledger that allows trustworthy audits. Blockchain technology supports these criteria as follows:

(1) *trust or control issues*: parties rely on a separate blockchain platform, as a neutral intermediary, to control the dataset and run the application logic. This platform is further considered "trustless": no trust is needed in a single provider to run the platform, but rather the proven consensus and ledger technologies for executing, validating and recording transactions and smart contracts. We point out that smart contracts, once deployed, cannot be influenced or tampered with (just like transactions).

(2) *shared and consistent dataset*: all involved parties have access to a historically consistent datastore[25] in terms of an immutable ledger; all parties are able to contribute data.

(3) *audit capabilities*: trustworthy audits are afforded by this immutable ledger: it records (a) all transactions that execute the smart contract, including any provided (possibly encrypted) input data, and (b) events emitted by the smart contract that reflect the outcomes.

---

[25] The need for privacy-sensitive identifiers can be met using anonymization techniques.

Below, we present multiple healthcare use cases and discuss their eligibility for blockchain deployment in light of these criteria.

## Clinical Trial Eligibility

When conducting clinical trials, there is a need to ensure adherence to regulatory requirements on research ethics and data privacy [7]; and avoid poor reproducibility of results when caused either by "honest mistakes" or fraud during the study [46]. These needs can be met in part by automating trial eligibility checking in a trustworthy and transparent way. Benchoufi et al. [4], Dai et al. [5], Zhuang et al. [6], and Benchoufi et al. [7] have studied the use of blockchain technology for trial eligibility. Regarding the suitability criteria:

(1) *trust or control issues*: no trust is required in study investigators to keep records, adhere to regulatory requirements, and avoid mistakes in eligibility (honest or otherwise). Instead, an appropriate smart contract can be deployed within the trustless blockchain setting. Clinical Trial Management Systems (CTMS) can also be utilized for this purpose, but they offer weaker safeguards as they are under single-party control [5].

(2) *shared and consistent dataset*: all involved parties, including the sponsors, patients, and investigators, will contribute ethical requirements, demographics, and health data.

(3) *audit capabilities*: regulators and trial sponsors as a matter of policy [46], or researchers in case of non-reproducibility, can conduct systematic audits of the blockchain ledger to assess regulatory adherence and rule out mistakes or fraud [4], [5].

## Health Screening Policies

In this public health use case, parties receive funding from the government to fund prevention and treatment programs based on screening outcomes. In prior work [8], we presented a diabetes risk screening use case where (1) regional healthcare organizations forward patient data to a region-wide platform, where an automated process determines diabetes risk, and (2) funding is subsequently distributed for diabetes prevention and treatment programs where they are needed most. Such programs include incentives for patients to keep their BMI below 26, walking 150 minutes or more per week, or equivalent exercises depending on their medical condition [47].

(1) *Trust or control issues*: healthcare providers will benefit from inflating diabetes risk factors among their population, may suspect such actions from other providers, or may perceive an underestimation of risk factors from the government. Instead, by deploying an appropriate smart contract on a blockchain-based system, no trust is required in any single party.

(2) *Shared and consistent dataset*: regional healthcare providers will contribute population data for the diabetes risk screening, which will be shared between all involved parties.

(3) *Audit capabilities*: in case of disagreement on screening outcomes, an audit of the ledger will reflect the given input population data (transactions) and the resulting screening outcomes (events), allowing the validation of screening outcomes.

## Multimorbidity Clinical Decision Support

For patients with multiple co-occurring illnesses (i.e., multimorbidity), appropriate treatments will often be decided by multiple different specialists. However, concomitant treatments for the same patient may adversely interact; e.g., a drug prescribed for one illness can nullify or potentiate the effect of another drug for a comorbid illness [48]. To avoid such conflicts, multimorbidity requires systematically sharing treatment decisions and, ideally, an automated process for detecting and resolving conflicts [49]. Van Woensel et al. suggest a loosely coupled approach for sharing treatment decisions [50]; when using blockchain as the shared treatment dataset in this approach, smart contracts can automate the detection and resolution of multimorbidity conflicts.

(1) *trust or control issues*: when multimorbidity conflicts occur, multiple treating physicians will often be responsible. No trust is required in fellow treating physicians, or their EHR systems, to check for conflicts or keep records of prescribed treatments; instead, an appropriate smart contract will check for conflicts in a trustless setting.

(2) *shared and consistent dataset*: the treating physicians will share treatment decisions on multimorbidity patients, allowing them to collaborate on conflict-free treatment plans.

(3) *audit capabilities*: in case of adverse health events in the patient, possibly caused by conflicting treatments, potential causes (and eventual liability) can be established by auditing the ledger for treatment decisions (transactions) and any conflicts that were detected and flagged by the smart contract (events).

## Health Insurance

In health insurance processes, such as claim handling, insurance companies decide which healthcare costs are reimbursed based on the patient's coverage and applicable policies. Automating these processes lowers the insurance's administrative costs and avoids human error. Karmakar et al. [9], Zhou et al. [10], and Chondrogiannis et al. [11] have explored the use of blockchain for medical insurance.

(1) *Trust or control issues*: patients, care providers, and insurers have competing interests and often a mutual distrust [9]: insurers will benefit from choosing the lowest-cost option; patients and care providers will benefit from maximizing reimbursement and health benefits. Deploying an appropriate smart contract on blockchain means no trust is required in a single party; instead, any choice on eligibility will be made and recorded in a trustless setting.

(2) *Shared and consistent dataset*: patients and care providers contribute information on the patient's health and treatment options; insurers contribute data on coverages and policies.

(3) *Audit capabilities*: in case of disagreement about claim decisions, audits of the ledger will reflect the provided health and coverage data (transactions), as well as resulting claim decisions (event), allowing the validation of the latter.

# Public vs. Private Consortium Blockchains

Regarding health insurance use cases, there exist two options for deployment: public or private consortium blockchains. As mentioned, in a public blockchain, anyone can join and perform transactions; cryptocurrency and decentralized finance transactions take place here. A private consortium chain is managed by multiple organizations and requires permission to join, and is typically used by banks and government organizations [43].

Deploying smart contracts on a public chain, such as the Ethereum mainnet, offers advantages including transparency, public verification of transactions, and greater interoperability. However, it also presents several challenges: a major consideration is privacy, as health insurance (and healthcare in general) involves sensitive personal information. The public nature of blockchain can pose privacy challenges, necessitating the encryption of transaction data and privacy protocols such as zero-knowledge proofs. Additionally, the health insurance industry is heavily regulated, and compliance with laws like HIPAA in the United States might be more challenging on a public chain, as a central entity cannot be held accountable in case of e.g., a data breach. Additionally, as mentioned, public networks can suffer from high transaction fees (gas costs) and may not scale well during periods of congestion, leading to unpredictable operating costs for health insurance applications.

A private consortium chain can offer greater control over who has access to the data, improving privacy for sensitive health information. Moreover, it can be optimized for higher transaction throughput at lower costs: the consensus mechanism can be tailored for a smaller number of known participants, and the rules of the network can be customized to suit the specific needs of the health insurance sector, e.g., compliance requirements and business logic. However, a private consortium blockchain has its own challenges. While private networks are more privacy-sensitive, they generally have fewer nodes, making them more susceptible to certain types of attacks or failures. Mitigating such attacks and failures necessitates increased trust among the participating nodes (such as healthcare partners), leading to a centralization risk that might not align with the "trustless" property of blockchain. At this point, we nevertheless foresee deploying health insurance smart contracts in a cost-optimized private consortium. This consortium would involve multiple healthcare organizations (e.g., insurance companies, care providers) that are familiar with each other, but nonetheless require a "trustless" setting and aim to leverage blockchain for that purpose.

## Using Logics-based Languages to Develop Smart Contracts

Idelberger et al. [13] reported in 2016 that logic-based languages have hardly been explored to implement smart contracts. To the best of our knowledge, this has not changed since then, aside from work by Choudhury et al. [19] and our prior work which this paper extends [8]. Idelberger et al. [13] showed that declarative, rule-based languages are more naturally suitable to represent legal clauses, i.e., they are easier to write and comprehend by domain experts such as jurists. Representing a license agreement in imperative procedural languages is far more involved, as it requires determining the ordering of imperative commands, the impact of triggers on the internal state, and propagating the state changes accordingly. A declarative rule-based program instead relies on an underlying rule engine for execution, without consideration of internal state or ordering. That said, the authors also outline an important technical challenge: reasoning algorithms have to be cheap, as per the economic rules of the blockchain—computational work and transactions expend the blockchain's cryptocurrency. With this in mind, the authors differentiate between *on-chain* and *off-chain* solutions for rule-based reasoning. We outline these two options below:

**Off-chain Solution**

A strictly off-chain solution involves deploying the rule engine as an oracle. The smart contract hereby emits "reasoning requests"; these are received by the oracle, which executes the rules and sends back inferences. This setup requires setting up a secure and scalable oracle, and an extra transaction to send back the inferences. Shukla et al. [27] developed the Read-Execute-Transact-Erase-Loop (RETEL) module, which similarly utilizes an oracle to outsource general application logic to off-chain Python scripts. Python scripts have a much larger set of libraries, and their computational work does not expend cryptocurrency. Nevertheless, this again requires a scalable and secure Python backend and an additional transaction. We only rely on oracles to retrieve data from remote IS (Figure 7); this is unavoidable in a distributed setting, where relevant data is stored by different parties.

**On-chain Solution**
A strictly on-chain option involves embedding an execution engine directly within the smart contract. Symboleo [51], PROForma [52], and GLEAN [22] implementations are based on Finite State Machine (FSM) execution semantics; Rasti et al. [53] manually implemented the Symboleo FSM in terms of base classes within smart contracts. These works are useful to implement decision logic limited to a particular domain, such as legal contracts. We target an approach for deploying general-purpose decision making, encoded by a semantic KG, on blockchain. To that end, a rule engine (reasoner) would have to be embedded, which tends to rely heavily on unbounded loop constructs, i.e., loops without obvious iteration limits. These are heavily discouraged on blockchain, as unbounded loops yield a-priori unknown execution costs—in other words, "on Ethereum, unlimited work is not an option"[26].

**Hybrid On-/Off-chain Solution:**
Alternatively, Idelberger et al. [13] mention the compilation of rule-based knowledge into a lower-level representation, to "increase the speed of inferential computation" within smart contracts. This is referred to as a hybrid on- and off-chain option, as compilation will occur off-chain and execution takes place on-chain. Our work belongs in this category; we compile a semantic KG, outfitted with N3 rules and an ontology, into the blockchain imperative programming language. Our evaluation shows the execution cost of our smart contract, and how it compares favorably. Nevertheless, our approach currently puts multiple restrictions on supported N3 rules; it remains an open question whether our approach would yield acceptable gas costs for more complex N3 rule sets. We revisit this in future work.

Choudhury et al. [19] similarly chose a compilation-based approach that generates smart contracts from rules written in SWRL. Their approach relies on apriori, manually written smart contract "templates": these implement the criteria involved, together with ADTs based on ontology classes, properties, and general constraints. Business logic rules written in SWRL [54] are parsed into an AST, which is then used to populate the placeholders within the pre-written template conditions (e.g., when checking for age, filling in the specific age constraint). The authors expect that smart contract templates will be applicable to all cases within a particular domain, such as clinical trial eligibility. Instead, we foresee our approach to be applicable to any domain with a domain standard encoded by an ontology.

As mentioned, our code generation-based approach essentially transforms a declarative interpretation (semantic KG) into an operational interpretation (imperative smart contracts).

---

[26] https://blog.b9lab.com/getting-loopy-with-solidity-1d51794622ad

Similarly, Verborgh and Arndt et al. [55] transform a declarative way of navigating hypermedia APIs into an operational interpretation, which executes HTTP requests and processes the data. In their declarative interpretation, a rule of the form { *<precondition>* } ⇒ { *<HTTP request> <postcondition>* } means the following: the existence of the precondition (e.g., link to a thumbnail) implies the existence of an HTTP request (e.g., GET) for the linked HTTP entity meeting the postcondition (e.g., thumbnail version). Their operational interpretation executes the GET request when the precondition is met, and then checks the result. This is similar to our approach of declaring the existence of HL7 FHIR resources (e.g., claims, coverages) at remote IS; in our operational interpretation, we issue a remote data request to the IS and then continue with the results.

## Code Generation for the Medical Domain

Past works have investigated imperative code generation for decision logic within the medical domain. Gietzelt et al. [56] present a syntax compiler for the Arden Syntax [57], which outputs Java Bytecode that can be executed on Java Virtual Machines (JVM). Arden Syntax is a domain-specific language for representing medical knowledge and can be used to express decision rules. This approach is specific to the Arden syntax and cannot be easily modified to cover a wider variety of clinical decision logic.

Porres et al. [58] present an automated approach to generate clinical decision support systems (CDSS) from clinical guidelines written as Unified Modelling Language (UML) state charts. The authors define a new set of UML attributes to better represent clinical guidelines in UML. To automatically generate code from the enriched UML document, the authors use MOFScript [59], an Eclipse plugin that supports control mechanisms and collection types. This approach is limited by the expressivity of UML state charts.

Wagholikar et al. [60] present experimental results from a CDSS for cervical cancer screening. The CDSS is composed of two rule bases, respectively generated from free-text and clinical guidelines. The former was generated by analyzing a large corpus of reports, and the latter was manually developed by experts based on national screening and management guidelines. The experiments found that the proposed system could output the optimal recommendations in 73 out of 74 cases. We outlined the issues with directly deploying rule engines on blockchain in the prior subsection.

## Blockchain for Health Insurance Scenarios

Due to our choice of a health insurance use case to present our work, we shortly summarize other works that utilize blockchain technology for health insurance.

Karmakar et al. [9] present ChainSure, a blockchain-based framework to help users find suitable health insurance policies, based on the Technique for Order of Preference by Similarity to Ideal Solution (TOPSIS). By choosing blockchain, insurance policies can be selected in a safe, transparent, and immutable way, while avoiding single-points-of-failure. Their focus lies on the identification of suitable insurance policies by implementing TOPSIS using smart contracts. MIStore by Zhou et al. [10] is a blockchain-based medical insurance storage system that ensures tamper-resistance and high credibility by employing a *(t, n)*-threshold protocol among all parties involved. Patient spending data is managed by a blockchain; servers enable insurance computations while preserving patient privacy. Hence, their focus lies on the secure storage and computation of insurance data.

Chondrogiannis et al. [11] introduce a Semantic Web and a blockchain-based compensation mechanism for research-related injuries (i.e., harm suffered by individuals participating in clinical studies). The authors deploy a "data check service" as an oracle that evaluates health contract terms against user-provided data. The service locates relevant insurance data, constructs an RDF graph, and checks the contract's conditions using an SPARQL ASK query. Hence, it relies on an off-chain oracle for computational work; we outlined the cons of this approach in the prior subsection. Our work generates imperative smart contract code directly from N3 logic, eliminating reliance on off-chain oracles for computational tasks.

## Conclusions

In a distributed Health 3.0 setting, decision making will be based on data from multiple parties. Here, blockchain is an excellent candidate for a decision-making platform, as it is able to act as a neutral intermediary; no party has to entrust another party with their privacy-sensitive data, nor trust them to be in charge of the decision making process. We outlined multiple healthcare scenarios where this "trustless" property of blockchain provides added value. At the same time, for decision making in a distributed environment, transmitted data will need to be semantically interoperable, i.e., structured using a domain standard (e.g., HL7 FHIR). Domain experts, who are most suitable for authoring decision logic as per these standards, typically lack the technical skills to write complex smart contracts within a distributed environment.

To that end, we presented a declarative approach for domain experts to encode high-level smart contract logic, in terms of a semantic KG (e.g., including remote data requirements). We deploy this KG on blockchain using a hybrid on-/off-chain solution: off-chain, a graph-based code generation pipeline compiles the KG into a smart contract, outfitted with remote data requests, which is subsequently deployed on-chain. We support multiple blockchain platforms by generating a bridge representation, which can be compiled into specific blockchain languages. Many design choices are informed by the economic rules of blockchain, i.e., where heavier computations result in higher execution costs. Firstly, our choice for off-chain code-generation avoids embedding a rule engine within a smart contract; these tend to rely heavily on unbounded loops, which yield a-priori unknown execution costs. Secondly, we instantiate ADTs within smart contracts based on terms referred to by the rule, instead of the entire OWL ontology (which can be huge). On Ethereum, smart contracts have a max. size of 24Kb; each byte of storage costs cryptocurrency as well. Finally, when possible, we implement multi-valued properties using a dictionary to avoid expensive loops.

We evaluated the generated smart contract code, visualizing their control flow and validating their correctness. We further measured the "gas" cost of their execution, which, while not trivial, compares favorably. We comprehensively positioned our work in the state of the art, and discussed the pros and cons of using a public vs. private consortium chain.

## Limitations and Future Work

A current limitation is the restricted set of N3 rules that are supported to express decision logic. We list these restrictions in the paper (section "Existential N3 Rules as Annotated Rule Graphs"). The feasibility of generating imperative code for more complex N3 rules, and the ensuing gas cost, remains an open question. For instance, a fixpoint-style algorithm could be used to implement recursive rulesets; however, this would essentially constitute an unbounded loop, which are heavily discouraged on blockchain. We aim to expand our approach towards more complex N3 rules in future work.

Secondly, Large Language Models (LLM) have been reported to be quite adept at code generation—possibly obviating the work from this paper. We previously studied the use of Large Language Models (LLM) to automatically generate N3 rules, and even smart contracts, from textual sources [61]. Although results were promising, we found that more complex scenarios—the ones for which automated support would be most useful—exceed the current ability of LLMs. Moreover, we further found that the generated code does not follow core principles of conceptual modeling, making the resulting code unsound; also, the LLM tends to overlook important details, while also hallucinating parts not found in the text. Nevertheless, we see LLMs as playing a role in supporting domain experts to author semantic KG (e.g., generating summaries of text, "skeleton" N3 rules). Moreover, future work involves further studying LLMs for smart contract code generation in this setting.

Thirdly, we only generated smart contract code for, and evaluated our approach on, a single blockchain (Ethereum). Future work involves applying our approach on other blockchains, such as HyperLedger.

Finally, we aim to expand our code generation approach in the following ways:
- Emit more fine-grained events that also report intermediary inferences and negative outcomes to the client (i.e., where a particular condition *does not* hold).
- For multi-valued property values, when a unique key for values is lacking but exists for groups of values (e.g., all values of type BMI), we aim to use a dictionary that is indexed on group keys and keeping arrays as values, to reduce the number of iterations needed.

## Acknowledgements

We would like to thank Dörthe Arndt for thoroughly reviewing the paper and pointing out inconsistencies with the RDF, OWL and N3 semantics.

## Conflicts of interest

None declared.

## Abbreviations

KG: Knowledge Graph
IS: Information Systems
ADT: Abstract Data Type

HL7 FHIR: Health Level 7 Fast Healthcare Interoperable Resources
API: Application Program Interface
REST: REpresentational State Transfer
CDSS: Clinical Decision Support Systems

# Appendix A: Rule Graphs

(Only highlighting rule heads in the graphs below for ease of illustration.)

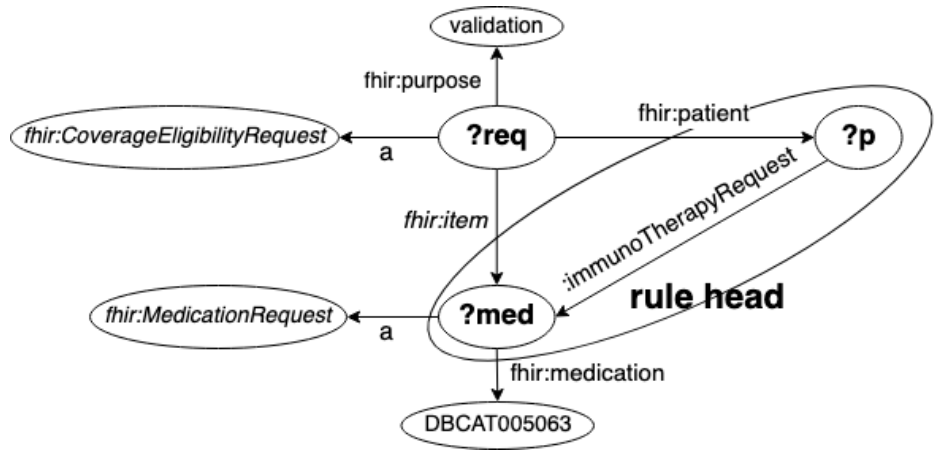

**Figure A.1**. Rule graph extracted from Code 1.

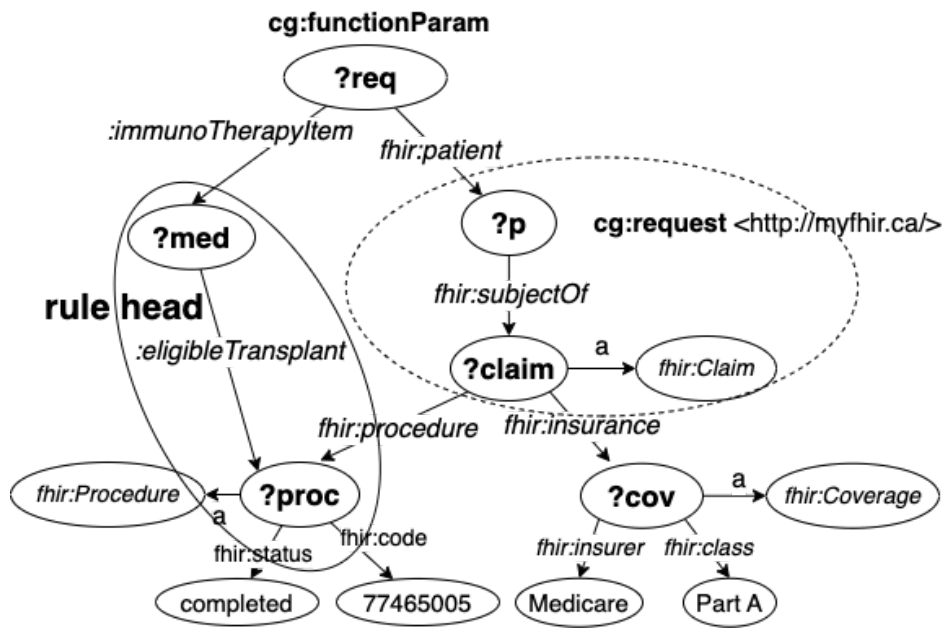

**Figure A.2**. Rule graph extracted from Code 2.

# Appendix B. Remote Data Requests in generateLogic

```
21.    for edge in node.outgoing do
s1.      if edge.tags includes 'cg:request' then
s2.        request ← RemoteRequest(location: edge.tags['request']) # remote url
s3.        remote_entity ← edge.target
s4.        request.select ∪ ← Parameter(prp: edge.original.term, value: path)
s5.        for redge in remote_entity.out do
s6.          rvalue ← redge.target.term
s7.          if redge.tags includes 'cg:request' then
s8.            if redge.term is rdf:type then
s9.              request.type ← rvalue # type value
s10.           else
s11.             request.select ∪<- Parameter(prp: redge.term, value: rvalue)
s12.       remote_result ← Variable()
s13.       block ∪ ← Assignment(remote_result, request)
s14.       path ← PropertyPath(remote_result)
s15.       outer_block ← block
s16.       conditions ← Conjunction()
s17.       block ← Block()
s18.       outer_block ∪ ← IfThen(conditions, block)
s19.       generateLogic(edge.target, edge, path, conditions, block)
s20.     else ... # continue with original code
```
**Algorithm B.1**. Remote data requests added to the generateLogic function.

For each outgoing edge (line 21), we check whether it is annotated with *cg:request* (line s1); if so, the edge will be used to select a subset of remote data. Line s2 creates a `RemoteRequest` instance, providing as location the request tag (e.g., myfhir.ca). E.g., for Code 3, we find selection parameters *<fhir:policyHolder> / <?p>* and *<fhir:status> / <active>*.

As shown in Fig. 3, to connect triple "`?cov fhir:policyHolder ?p`" to the *?req* function parameter, the inverse property *policyHolderOf* was used. As the algorithm starts from the request parameter *?req*, the edge *policyHolderOf* will be the first edge found annotated with *cg:request*. At this point, the edge's target *?cov* is "remote entity", i.e., entity about which remote data is retrieved (line s2), and the property path will be `req.patient`:

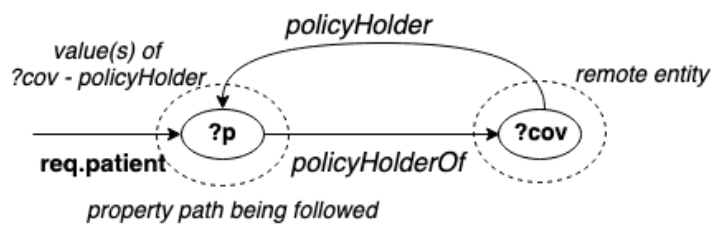

We observe that the property path `req.patient` points towards values of the original *policyHolder* property. Hence, line s4 adds as select *Parameter* the current edge's "original" term (*policyHolder*) as property and the current path (`req.patient`) as value.

This pattern occurs for any case that follows the structure outlined in Remote Data Requests, and *<value i>* of one of the property-value pairs[27] is a variable:
```
{  ?x a <type> ; <property 1> <value 1> ... ; <property n> <value n> }
```

For our example:
```
{ ?cov a fhir:Coverage ; ?cov fhir:policyHolder ?p ; fhir:status 'active' }
```

When attempting to create a coherent rule graph, the parser will connect the *<value i>* variable to the function parameter using an inverse property. E.g., Figure A.2 similarly connects triple "`?claim fhir:subject ?p`" to *?req* using the inverse *subjectOf*. Hence, the `req.patient` property path will similarly provide values for the original *subject* property.

On lines s5-s11, the function iterates over all outgoing edges of the remote entity; in case an edge is similarly tagged with *cg:request* (line s7), then it will be used as either the type of data to be retrieved (lines s8-s9) or as an additional parameter (lines s10-s11).

Note that any subsequent path will rely on the retrieved data; e.g., conditions that check the *fhir:insurer* and *fhir:class* (Fig. 3) of the retrieved *?cov* values. To that end, lines s12-s13 instantiate an `Assignment` of the remote request results to a *remote_result* variable, and line s14 instantiates *a new property path starting from remote_result*. To collect conditions, operations and assignments on the returned results, lines s15-s18 instantiate a new `IfThen` with a new `Block` and `Conditions`. Finally, the *generateLogic* function is then recursively called for the remote entity (line s19).

---

[27] In case multiple <value>s are variables, then they will be resolved using unification.

# Appendix C. Multi-Valued Properties in generateLogic

```
...
2.  function generateLogic(node, from, path, conditions, block)
...
20.   # (end if)
n1.   if last_property(path) is multi-valued then
n2.     if any edge in node.out where edge.term is rdf:type then
n3.       # (1) index values on unique key
n4.       nary_iteration ← false
n5.       last_property(path).use_dictionary ← true
n6.       key ← edge.target.term # use value of rdf:type as unique key
n7.       last_property(path).key  ← key
n8.     else
n9.       # (2) iterate over values
n10      nary_iteration ← true
n11.      outer_conditions ← conditions
n12.      outer_block ← block
n13.      original_path ← path
n14.      loop_var ← Variable() # start paths from loop variable
n15.      path ← PropertyPath(loop_var)
n16.      # collect any nested conditions & statements here
n17.      conditions ← Conjunction();
n18.      block ← Block()
21.   for edge in node.outgoing do
      ...
n19.  if nary_iteration then
n20.    if conditions is not empty then
n21.      if block is not empty then
n22.        # both conditions and assignments were provided
n23.        outer_block U ← Iteration(original_path, loop_var,
n24.           IfThen(conditions, block))
n25.      else
n26.        # only conditions were provided
n27.        outer_conditions U ← QualifiedCondition(EXISTENTIAL, outer_path,
n28.           loop_var, conditions)
n29.    else if block is not empty then
n30.      # only assignments were provided
n31.      outer_block U ← Iteration(outer_path, loop_var, block)
```

**Algorithm C.1**. Multi-valued properties added to the generateLogic function.

In case the latest property in the *path* is multi-valued (line n1), the current *node* parameter will represent its value. For this node, we check whether any outgoing edge is the type property (line n2): if so (lines n3-n7), we will utilize a dictionary to keep the property's values and use the provided type value as key to access the dictionary (line n6-n7). See Code 9 for an example of the generated application logic.

Otherwise, we will need to iterate over the multi-valued property values (lines n8-n18). We assign the "original" *conditions*, *block* and *path* parameters to temporary variables (lines n11-n13). Next, we create a new *PropertyPath*, starting from a new "loop" *Variable* (lines n14-n15): any property path in the iteration body will start from this loop variable (e.g., "item" variable in Code 11). We then create a new *Conjunction* and *Block* to capture conditions and assignments to be nested in the iteration body (lines n17-n18). Then, as before, the function proceeds by recursively calling the *generateLogic* function, passing the (possibly newly created) *conditions* and *block* as parameters.

Afterwards, in case an iteration was required (line n19), we process the collected *conditions* and *block* that will be nested within the iteration. At execution time, in each iteration, a value from the *outer_path* will be assigned to the *loop_var*, and the iteration body is executed. Regarding the iteration body, we consider the following cases:

(**1**) Both conditions and assignments were given (lines n21-n24). An *Iteration* instance is added to the *outer_block*, where a nested *IfThen*, with the collected *conditions* and *body*, will serve as the iteration body. E.g., see Code 10.

(**2**) Only conditions were given (lines n25-n27). A *QualifiedCondition* with an existential qualifier and collected *conditions* is added to *outer_conditions*. E.g., see Code 12 (line 3).

(**3**) Only assignments were given (lines n28-n30). Similar to (1), an *Iteration* instance is added to the *outer_block*, where the collected *block* will serve as the iteration body.

# Appendix D. Operations in generateLogic

We use the unification mechanism to support mathematical (and possibly) other operations and their nesting. Below, we show example N3 code to calculate a patient's BMI:

```
1.   ?patient :height ?h ; :weight ?w .
2.   (?h 2) math:exponentiation ?h_exp .
3.   (?w ?h_exp) math:quotient ?bmi .
```

**Code 7**. Example BMI calculation using N3 operations.

N3 operations are typically expressed as statements with as subject a list of operands (e.g., *?h*, 2), a predicate indicating the operator (e.g., *math:exponentiation*), and an object variable (e.g., *?h_exp*, *?bmi*) that captures the result. To support operations, we use the unification variable mapping to further keep any operations associated with variables. For the above example, the variable map would keep the following property paths and operations:

```
?h     ⇒ patient.height
?w     ⇒ patient.weight
?h_exp ⇒ ( ?h 2 ) math:exponentiation
?bmi   ⇒ ( ?w ?h_exp) math:quotient
```

Using the variable mapping, the *generateLogic* function will attempt to recursively resolve all operands of the operations. In a nutshell, operations will have the resolved variables as operands (e.g., property paths, or variables from other operations), and an `Assignment` is created between the `Operation` and the "result" `Variable`. For the example above, this yields the following pseudocode:

```
h_exp = p.height ^ 2;
bmi = p.weight / h_exp;
```

We assume that all triples with subject collections (e.g., Code 7, lines 2-3) represent operations. From the rule graph, all edges with a collection as source node are separately collected (not shown) into the *operations* array; subsequently, the *generateOperations* function is called:

```
1.  function generateOperations(operations):
2.    while true do
3.      num1 ← operations.length
4.      for each edge in operations do
5.        operands ← edge.source
6.        if generateOperation(edge) then
7.          operations \ ← edge  # remove edge from operations set
8.      num2 ← operations.length
9.      if num2 == 0 then return
10.     if num1 == num2 then error  # no operations could be resolved
```

**Algorithm D.1**. Generate Operations Function Pseudocode.

This function implements a fixpoint algorithm to resolve operations. This is required as operations may refer to the results of other operations in any ordering within the rule. On lines 4-7, for each graph edge in *operations*, the function calls the *generateOperation* function (see below). If successful, this means all of the operation's operands were resolved, and the edge is removed from the collection. If no *operation* edges remain, then the algorithm returns (line 9); else, if no additional operations were resolved in a given iteration (line 10), an error is thrown (we reached an incorrect fixpoint). Else, we continue with the next iteration.

```
1.  global: preamble:Block
2.  function generateOperation(edge):
3.    operands ← edge.source # subject is collection of operands
4.    for each operand in operands do
5.      if operand is variable then
6.        if operand in var_map then
7.          resolved U ← operand
8.        else # variable operand not yet resolved
9.          return false
10.     else # concrete operand
11.       resolved U ← operand
12.   operator ← edge.term # predicate is operator
13.   variable ← Variable(edge.target) # object is variable
14.   operation ← Operation(resolved, operator) # resolved = operation arguments
15.   var_map[edge.target] ← operation
16.   preamble U ← Assignment(variable, operation)
```
**Algorithm D.2**. Generate Operation Function Pseudocode.

On lines 3-11, the function attempts to resolve the *operands* of the given operation: if the *operand* is a variable, it will attempt to retrieve its unification from the variable map; if any *operand* cannot be resolved, the function returns false (lines 5-9). Subsequently, on lines 12-14, the function will construct an `Operation` given the resolved *operands* and the *operator*. The variable is mapped to the newly created operation (line 15); and an `Assignment` between the *variable* and the *operation* is added to the "preamble" of the `IfThen` construct (line 16), i.e., it will be added as a separate statement before the "if" clause. As a result, any subsequent comparison or assignment (or constructor parameter) may refer to the variable.

# Appendix E. Evaluation Results

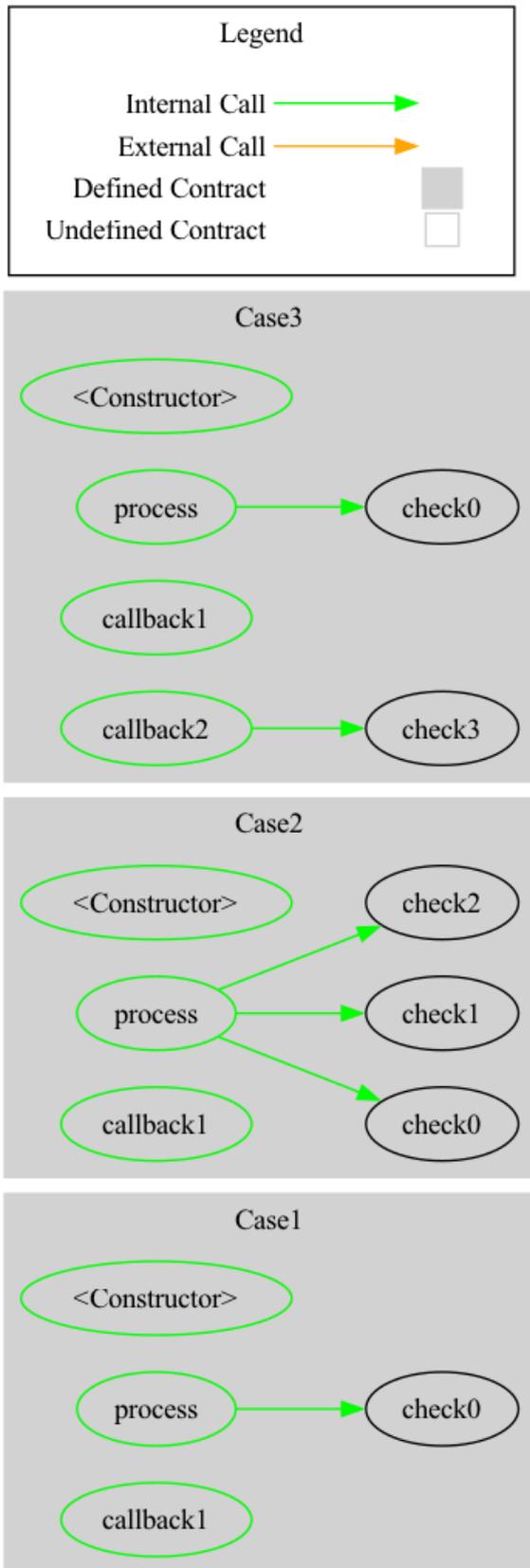

**Figure 8**. Internal / external function calls of the 3 smart contracts captured by Surya [40].

**Structs:**
1. CoverageEligibilityRequest represents a request for coverage eligibility, containing details such as the purpose, a list of medication requests, specific information about immunotherapy, patient details, and a response.
2. MedicationRequest holds information about requested medications and eligibility for a specific procedure.
3. Medication represents a medication with a unique concept identifier.
4. Procedure contains details about a medical procedure, including its status and a code.
5. Patient stores patient information, including ID, claims, and coverage details.
6. Claim contains details about a procedure and insurance.
7. Insurance holds information about an insurer and the class of insurance.
8. Insurer contains details about the insurer, such as type (e.g., Medicare).
9. Coverage represents coverage information, including details about the insurer, class, and status.
10. CoverageEligibilityResponse stores the outcome of a coverage eligibility request.
11. RequestData is used for oracle requests, containing the coverage eligibility request and additional parameters.

**Enums:**
1. Insurers: Defines types of insurers (e.g., Medicare).
2. Classes: Defines classes of insurance (e.g., PartA, PartB, PartD).

**Events:**
1. ContractResponse: Emitted when a coverage eligibility response is determined.
2. OracleRequest: Emitted for requesting data from an external oracle.

**Functions:**
1. process() is the main function for handling a coverage eligibility request. It checks certain conditions and, if criteria are met, then emits an "OracleRequest" event.
2. callback1() and callback2() are the functions designed to be called in response to oracle data. They process the received data and may trigger state changes or further actions, such as emitting a ContractResponse event.